\documentclass[twocolumn,aps,prl,preprint,superscriptaddress]{revtex4-2}
\usepackage{graphicx}
\usepackage{geometry}
\usepackage{setspace}
\usepackage{ragged2e}
\usepackage{lipsum}
\usepackage{amsmath}
\usepackage{float}
\renewcommand{\raggedright}{\leftskip=0pt \rightskip=0pt plus 0cm}
\usepackage[format=plain,margin=0pt,justification=raggedright,font=scriptsize,labelfont=bf]{caption}

\begin{document}
\preprint{}

\parskip=0pt 
\title{Two-photon MINFLUX with doubled localization precision}


\author{Kun Zhao}
\affiliation{Department of Biomedical Engineering, College of Future Technology, Peking University, Beijing 100871, China.}
\affiliation{Wallace H Coulter Department of Biomedical Engineering, 
	Georgia Institute of Technology and Emory University, Atlanta 30332, Georgia, USA.}
\author{Xinzhu Xu}
\affiliation{Department of Biomedical Engineering, College of Future Technology, Peking University, Beijing 100871, China.}
\affiliation{Wallace H Coulter Department of Biomedical Engineering, 
	Georgia Institute of Technology and Emory University, Atlanta 30332, Georgia, USA.}
\author{Wei Ren}
\affiliation{Department of Biomedical Engineering, College of Future Technology, Peking University, Beijing 100871, China.}
\author{Peng Xi}
\email[Corresponding Author: ]{xipeng@pku.edu.cn}
\affiliation{Department of Biomedical Engineering, College of Future Technology, Peking University, Beijing 100871, China.}
\thanks{Peng XI}
\email[]{xipeng@pku.edu.cn}

\altaffiliation{Department of Biomedical Engineering, College of Future Technology, Peking University, Beijing 100871, China.}

%
\begin{abstract}
\begin{footnotesize}
Achieving localization with molecular precision has been of great interest for extending fluorescence microscopy to nanoscopy. MINFLUX pioneers this transition through point spread function (PSF) engineering, yet its performance is primarily limited by the signal-to-background ratio. Here we demonstrate that applying two-photon excitation to MINFLUX would double its localization precision through PSF engineering by nonlinear effect. Cramér-Rao Bound (CRB) is studied as the maximum localization precision, and CRB of two-photon MINFLUX is halved compared to single-photon MINFLUX in all three dimensions. Meanwhile, in order to achieve same localization precision with single-photon MINFLUX, two-photon MINFLUX requires only 1/4 of fluorescence photons, contributing to a possible 4-fold temporal resolution. Benefitted from two-photon excitation, registration-free multicolor nanoscopy and ultrafast 
color tracking can be achieved. Localization precision of MINFLUX can also be doubled using excitation with second-order Laguerre-Gaussian beams but would suffer from high fluorescence background compared to single-photon and two-photon first-order MINFLUX.
\end{footnotesize}
\end{abstract}


\maketitle

\begin{footnotesize}
Super-resolution microscopy takes our vision from the conventional 200 nm diffraction limit down to 20 nm regime. MINFLUX further extends the resolution to sub-10 nm \cite{balzarotti2017nanometer,eilers2018minflux,pape2020multicolor,gwosch2020minflux} , through combining the coordinate-targeted STED donut \cite{hell1994breaking} and the coordinate-stochastic single-molecule localization microscopy (SMLM) \cite{betzig2006imaging,2006Stochastic}. MINFLUX can achieve such high resolution because of its 3D localization precision of single-digit nanometer. By engineering the point spread functions of the microscope, many techniques were introduced to break the precision confinement of localization \cite{hell1994confocal,gustafsson1999i5m,bewersdorf2006comparison,pavani2009three,jia2014isotropic,backlund2018fundamental}. In MINFLUX \cite{balzarotti2017nanometer,eilers2018minflux,pape2020multicolor,gwosch2020minflux}, the excitation PSF is engineered to first-order Laguerre-Gaussian beam. Using the intensity minimum of coordinate-targeted confocal excitation, MINFLUX reduce much of the required photons, as intensity minimum has much higher contrast of intensity than intensity maximum of Gaussian excitation, thus is less prone to Poissonian noise \cite{balzarotti2017nanometer,xiao2017flipping}. Yet, full potential of MINFLUX has not been reached regarding acquisition time and signal-to-background ratio.

Multiphoton microscopy features with a nonlinear dependence of fluorescence to excitation, hence bear potential to further increase the spatio-temporal resolution of MINFLUX. For example, in two-photon microscopy, intensity of fluorescence is proportional to the square of excitation intensity \cite{peticolas1963double,sheppard1990image,denk1990two,denk2006multi}. The contrast at intensity minima could be enhanced by the square dependence. Intuitively, together with that two-photon excitation decrease fluorescence background, employing two-photon excitation to MINFLUX can further improve localization precision. 

Here we give the theoretical framework of two-photon  MINFLUX (2p-MINFLUX). We show explicitly that the square of fluorescence intensity results in 2-fold increase in maximum localization precision compared to single-photon MINFLUX. This suggests, more importantly, that only 1/4 photons are needed for achieving the same localization precision. Further, taking advantage of spectral overlapping of absorption for different fluorophore  \cite{bestvater2002two,mutze2012excitation,velasco2015absolute,xu2015multiphoton}, two-photon MINFLUX  would provide registration-free multicolor localizations through emission spectral separation, which paves new avenue for simultaneous ultrafast single particle tracking of fluorophores of different spectra.

For a quantitative illustration, suppose ideal donut excitation with zero-center and background-free condition with only Poissonian noise. Representing the estimations with confidence intervals [Fig. 1(a)], we could see the width of confidence interval for two-photon fluorescence localization is 1/2 of width for single-photon fluorescence, showing better localization precision by two-photon donut excitation.

To explicitly evaluate the improvement, Cramér-Rao Bound (CRB) is calculated for maximum localization precision of 2p-MINFLUX \cite{balzarotti2017nanometer}. Modification is made on the Poissonian mean $\lambda$ for two-photon fluorescence considering nonlinear effect:
\begin{subequations}
\begin{equation}
\lambda_{1p}\left(\vec{r}_{f}\right)=f_{1} I_{1p}\left(\vec{r}_{f}\right)\label{1a}
\end{equation}
\begin{equation}
\lambda_{2p}\left(\vec{r}_{f}\right)=f_{2} I_{2p}^{2}\left(\vec{r}_{f}\right)\label{1b}
\end{equation}
\end{subequations}
where \begin{math}\vec{r}_{f}\end{math} is the fluorophore position, $f_{1}$ and $f_{2}$ stand for, for simplicity, factors corresponding to absorption cross-section of fluorophore, quantum yield and collection efficiency of the system, and $I_{\rm{1p}}$ and $I_{\rm{2p}}$ are point spread functions (PSF) of donut excitations. All other parameters in the model are kept unchanged.

CRB is expressed in an intricate general form of $\frac{\partial{\lambda}}{\partial x}$ and $\frac{\partial {\lambda}}{\partial y}$. At the origin where localization precision is highest (\textit{i.e.}, minimum CRB), CRB can be expressed explicitly as:
{\setlength\abovedisplayskip{0.3cm}
	\setlength\belowdisplayskip{0.3cm}
	\begin{subequations}
	\begin{equation}
	\begin{split}
	\label{2a}
CRB_{1p}(\overrightarrow{0})&=\frac{L}{2\sqrt{2N}} \frac{s}{1-\frac{L^{2} \ln 2}{fwhm_{1p}^{2}}}
	\end{split}
	\end{equation}
	\begin{equation}
\begin{split}
\label{2b}
CRB_{2p}(\overrightarrow{0})&=\frac{L}{4\sqrt{2N}} \frac{s}{1-\frac{L^{2} \ln 2}{fwhm_{2p}^{2}}}
	\end{split}
\end{equation}
\end{subequations}
}where \textit{L} is diameter of targeted coordinate pattern (TCP) circle, \textit{N} is number of detected photons, \textit{fwhm} is the full width at half maximum of the excitation PSF, and $s=\sqrt{\left(\frac{1}{SBR}+1\right)\left(\frac{3}{4SBR}+1\right)}$, where \textit{SBR} is signal-to-background ratio. Since $ \textit{L} \ll \textit{fwhm$_{1p}$} < \textit{fwhm$_{2p}$}$, we can obtain precision increase slightly larger than two-fold:
{\setlength\abovedisplayskip{0.3cm}
	\setlength\belowdisplayskip{0.3cm}
 \begin{equation}
 CRB_{2p}(\overrightarrow{0})\leq \frac{1}{2} CRB_{1p}(\overrightarrow{0})\label{3}
 \end{equation}
}

The reason for halving of CRB lies in that a factor of 2 appears when the square of $I_{2p}$ is differentiated:
\begin{subequations}
 \begin{equation}
\frac{\partial \lambda_{1p}}{\partial x}=\frac{\partial\left(f_{1} I_{1p}\right)}{\partial x}=f_{1} \frac{\partial I_{1p}}{\partial x}\label{4a}
\end{equation}
 \begin{equation}
\frac{\partial \lambda_{2p}}{\partial x}=\frac{\partial\left(f_{2} I_{2p}^2\right)}{\partial x}=2f_{2}I_{2p} \frac{\partial I_{2p}}{\partial x}\label{4b}
\end{equation}
\end{subequations}
and that in CRB formula (see Eq.(S26) in \cite{balzarotti2017nanometer}), the denominator has one more power of the above partial derivatives than the nominator, resulting in an additional factor of 2 in the denominator.

Maximum localization precision is compared for single-photon and two-photon MINFLUX with respect to \textit{N} and \textit{L} [Fig. 1(b)]. In our simulation, wavelengths are set to 647 nm and 800 nm for single-photon and two-photon excitation respectively. Achievable \textit{SBR} for 2p-MINFLUX should be measured in experiments; experimental \textit{SBR} of 1p-MINFLUX is used in our calculation. 2p-MINFLUX, utilizing 1/4 photons compared with 1p-MINFLUX, possesses same or even higher localization precisions as \textit{L} increases. For \textit{L} = 50 nm and \textit{N} = 100, $CRB_{2p}$ and $CRB_{1p}$ are 0.97 nm and 1.96 nm respectively, with CRB enhancement ratio $R_{CRB}$ = ${CRB_{1p}}$/${CRB_{2p}}$=2.02.

\begin{figure}[t]
	\includegraphics[trim=20 0 10 0, scale=0.9]{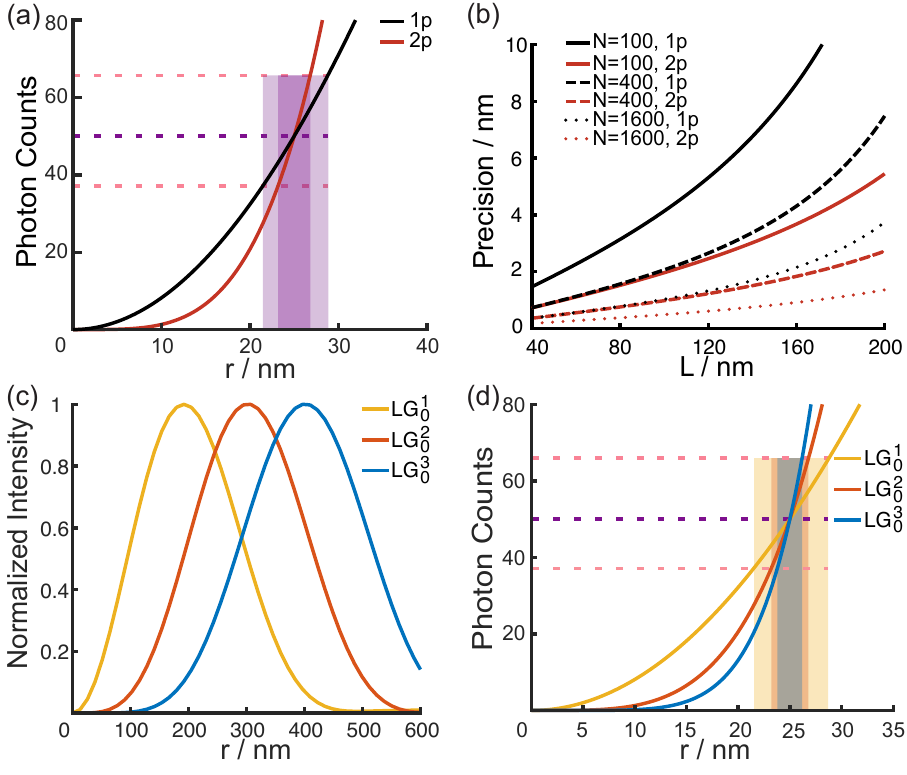}%
	\caption{Localization precision of 2p-MINFLUX. (a) Intuitive understanding of precision enhancement of 2p-MINFLUX. Length of 95\% confidence interval of localized position is halved for two-photon donut. Assume Poissonian means for single-photon and two-photon fluorescence at \textit{r} = 25 nm are both 50, and assume detected photon is 50 as well. (b) Comparison of the maximum localization precision between single-photon and two-photon MINFLUX with respect to \textit{N} and \textit{L}. (c) Intensity of tightly focused Laguerre-Gaussian beams at focal plane. Electric field is simulated with $\lambda$ = 647 nm, $\alpha$ = 69.0°, \textit{f}/\textit{w} = 0.58, and \textit{fwhm} is calculated with NA = 1.4. (d) Intuitive understanding of precision enhancement with higher-order Laguerre-Gaussian beams. Assumptions are made same as (a).}
\end{figure}

We study as well the possibility of applying higher-order Laguerre-Gaussian beam to MINFLUX \cite{wolf1959electromagnetic,richards1959electromagnetic,youngworth2000focusing,zhan2006properties,zhao2007spin,xie2013analytical}. Tightly focused Laguerre-Gaussian beams $\mathrm{{LG}_{0}^{\textit{l}}}$ are calculated at the focal plane for 3 orders with \textit{l}=1,2 and 3 [Fig. 1(c)] \cite{wolf1959electromagnetic,richards1959electromagnetic,youngworth2000focusing,zhan2006properties,zhao2007spin,xie2013analytical}. For \textit{r}$\ll$ \textit{fwhm}, $\mathrm{{LG}_{0}^{2}}$ and $\mathrm{{LG}_{0}^{3}}$ can be approximated by the square and cube of $\mathrm{{LG}_{0}^{1}}$, which would translate into 2-fold and 3-fold increase of localization precision [Fig. 1(d)], similar to the effect of two-photon or three-photon excitation. However, excitation intensity decreases sharply for \textit{r}$\ll$ \textit{fwhm} for $\mathrm{{LG}_{0}^{2}}$ and $\mathrm{{LG}_{0}^{3}}$ compared to $\mathrm{{LG}_{0}^{1}}$. This raises concern for the attainable \textit{SBR}, since in order to gain enough fluorescence signal, the total excitation power should be increased, which may lead to higher fluorescence background.

CRB across 2D xy-plane for one-photon and two-photon MINFLUX is compared for \textit{L} = 50 nm [Fig. 2(a)-(c)].  Of most interest is only CRB in a small center region of TCP circle instead of whole xy-plane, since, in iterative MINFLUX \cite{gwosch2020minflux}, a previous round already localizes the fluorophore to a confined region. A center region with diameter of 10 nm has  $R_{CRB}$ \textgreater 1.75 [Fig. 2(c)-(d)].

\begin{figure}[b]
	\flushleft
	\includegraphics[scale=0.9,trim=0 -15 0 0]{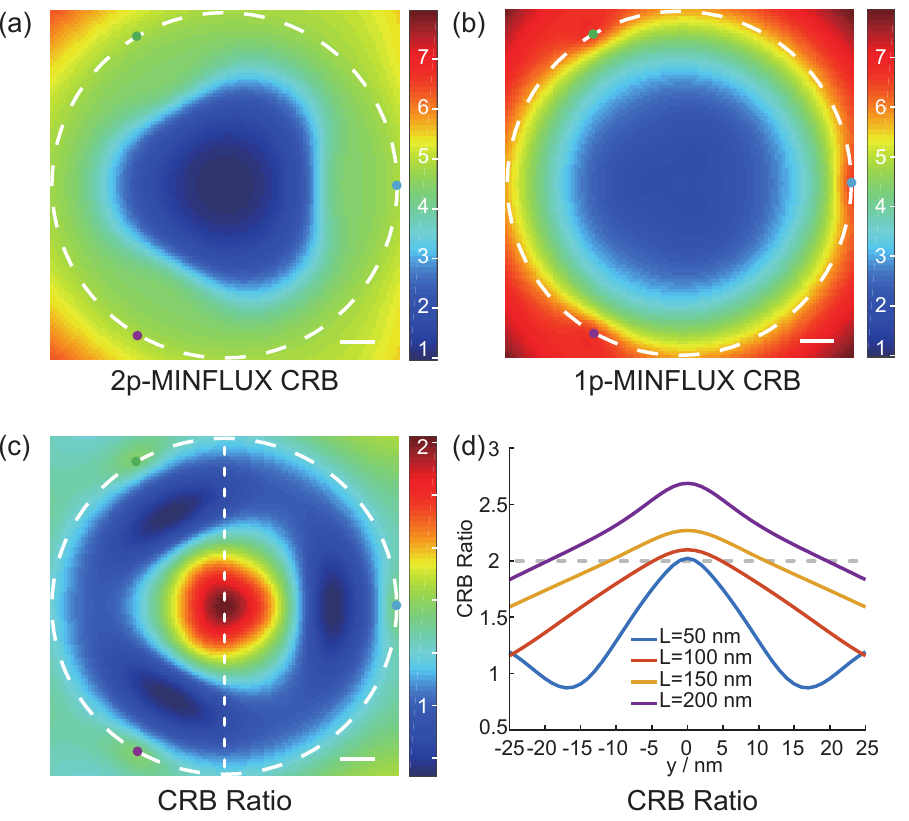}%
	\caption{Comparison of CRB of 2p-MINFLUX versus 1p-MINFLUX in xy-plane. (a,b) Comparison of CRB across xy-plane of (a) 2p-MINFLUX and (b) 1p-MINFLUX. Colormaps are rescaled into same dynamic range. CRB is calculated with \textit{N} = 100, \textit{L} = 50 nm, and \textit{SBR} = 12. (c) Ratio of CRB of 1p-MINFLUX to 2p-MINFLUX in (a,b). Dotted line (x=0) represents direction for the 1D profiles in (d). (d) 1D cut profile of CRB ratio for different \textit{L}. Scale bar: 5 nm.}
\end{figure}

CRB enhancement ratio $R_{CRB}$ across xy-plane is compared for a series of \textit{L} = \{50, 100, 150, 200\} nm using representative 1D profiles [Fig. 2(d)]. $R_{CRB}$ increases as \textit{L} increases, which could potentially be exploited for further CRB enhancement in initial rounds with larger \textit{L} in iterative MINFLUX \cite{gwosch2020minflux}.

The enhancement of z-localization precision is similar to xy-localization. 3D-donut is modeled simply as a quadratic function \cite{gwosch2020minflux}. The highest precision at the origin also increases by 2-fold:
{\setlength\abovedisplayskip{0.3cm}
	\setlength\belowdisplayskip{0.3cm}
	\begin{subequations}
\begin{equation}
CRBz_{1p}(\overrightarrow{0})=\frac{L}{4\sqrt{N}}\left(\frac{1}{SBR}+1\right)\label{5a}
\end{equation}
\begin{equation}
CRBz_{2p}(\overrightarrow{0})=\frac{L}{8\sqrt{N}}\left(\frac{1}{SBR}+1\right)\label{5b}
\end{equation}
\end{subequations}
}

Further, to demonstrate the axial enhancement, CRB in z-axis is compared between 2p-MINFLUX and 1p-MINFLUX for different \textit{L} [Fig. 3(a)-(b)]. Precision of 2p-MINFLUX decreases rapidly with \textit{z}, likely because of singularities occurred at positions of TCP (z = \textit{-L}/2 and \textit{L}/2). CRB enhancement ratio decreases along \textit{z}, similar to localizations in xy-plane.

\begin{figure}[b]
	\flushleft
	\includegraphics[scale=0.9,trim=-5 -3 20 0]{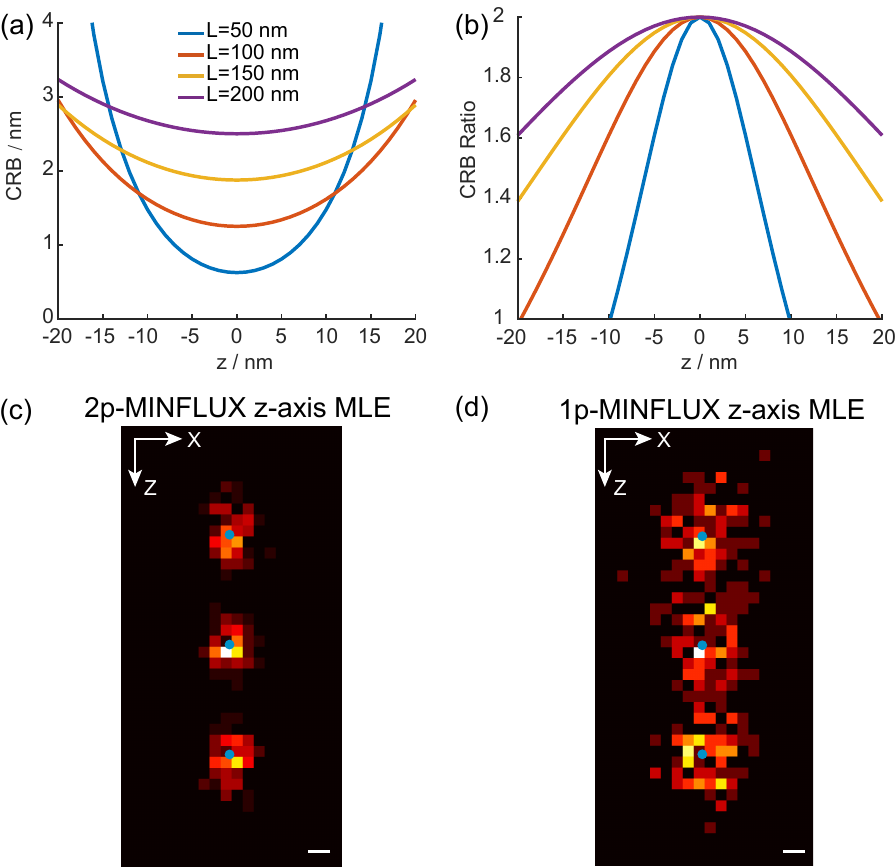}%
	\caption{Performance of 2p-MINFLUX in z-axis. (a) z-axis CRB of 2p-MINFLUX for different \textit{L}. Calculation is conducted with \textit{N} = 100, \textit{L} = 50 nm, and \textit{SBR} = 12 for all subfigures. (b) Ratio of z-axis CRB of 1p-MINFLUX to 2p-MINFLUX. (c-d) Simulation of z-axis localization of 2p-MINFLUX and 1p-MINFLUX. Distances between each adjacent dyes are 5 nm. For visualization, \textit{x} positions are generated randomly with normal random number. Pixel size is 0.5 nm. Scale bar: (c-d) 1 nm.}
\end{figure}

Using maximum likelihood estimation (MLE), z-axis localization, as a 1D problem, can be solved analytically for 2p-MINFLUX as well:
{\setlength\abovedisplayskip{0.3cm}
		\setlength\belowdisplayskip{0.3cm}
\begin{subequations}
\begin{equation}
z_{1p}^{(MLE)}=-\frac{L}{2}+\frac{L}{1+\sqrt[2]{\frac{n_{1}}{n_{0}}}}\label{6a}
\end{equation}
\begin{equation}
z_{2p}^{(MLE)}=-\frac{L}{2}+\frac{L}{1+\sqrt[4]{\frac{n_{1}}{n_{0}}}}\label{6b}
\end{equation}
\end{subequations}
}where $\textit{n}_{0}$ and $\textit{n}_{1}$ are number of photons detected with $I\left(-\frac{L}{2}\right)$ and $I\left(\frac{L}{2}\right)$. Imaginary roots and a real root outside of $\left(-\frac{L}{2}, \frac{L}{2}\right)$ are neglected for 2p-MINFLUX. In simulation, the above estimators agree with ground truth, and enhancement of localization precision can be seen clearly [Fig. 3(c)-(d)].

\begin{figure}[b]
	\flushleft
	\includegraphics[scale=1.2,trim=-10 0 0 0]{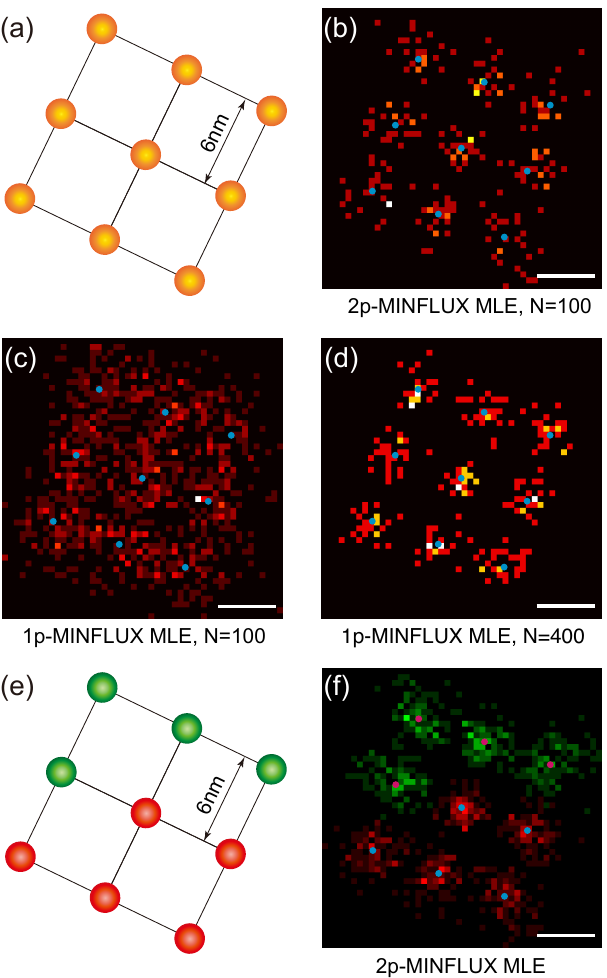}%
	\caption{Simulation results of 2p-MINFLUX on dyes grids. (a) Configuration of a simulated single-color dye-labelled square DNA origami. Distances between each adjacent dyes are 6 nm. (b-d) Simulation on the above origami of (b) 2p-MINFLUX and (c,d) 1p-MINFLUX for comparison. For each dye, simulation of localization was repeated for (b,d) 25 and (c) 100 times respectively, with \textit{N} = 100 (b,c) or 400 (d), \textit{L} = 50 nm, and \textit{SBR} = 12. Pixel size is 0.5 nm. (e) Configuration of a simulated dual-color dye-labelled origami for simulation of dual-color 2p-MINFLUX. Angle and distance parameters are same with origami in (a). (f) Simulation of 2p-MINFLUX on the above origami utilizing MLE estimator. Simulation was repeated for 100 times, with \textit{N} = 100, \textit{L} = 50 nm, and \textit{SBR} = 12. Scale bar: (b-d, and f) 5 nm.}

\end{figure}

In addition, we simulated MINFLUX imaging in xy-plane for \textit{L} = 50 nm with numerically solved MLE estimation [Fig. (4)]. For photon number \textit{N} as low as 100, 2p-MINFLUX can already achieve 6 nm resolution, which is not feasible in 1p-MINFLUX [Fig. 4(b)-(c)]. In addition, compared with 1p-MINFLUX with \textit{N}=400 [Fig. 4(d)], 2p-MINFLUX with \textit{N}=100 achieved similar localization distributions at the origin, confirming the capability of 2p-MINFLUX to reduce number of photons required. This 4-fold decrease of required photons would potentially result in 4-times faster localization speed, and 4-times more localizations per fluorophore due to limited total fluorescence photons, beneficial for MINFLUX as both a super-resolution microscopy technique and also a single-particle tracking technique.

We believe that two-photon MINFLUX would be capable of multicolor localizations [Fig. 4(e)-(f)]. Because of the spectral overlapping of two-photon absorption peaks, it is possible to excite multiple fluorophores simultaneously. Although it is not a must to use a single-wavelength excitation for multicolor two-photon microscopy, a single-wavelength two-photon excitation is beneficial for multicolor MINFLUX. 2p-MINFLUX would be free of registration of different color channels, as they are excited with the same donut coordinates. Hence, this enables simultaneous registration-free, multicolor, ultrafast  MINFLUX tracking, which is crucial for study of molecular interaction.

In conclusion, MINFLUX can be enhanced with multiphoton excitation process, with 2-fold increase of localization precision or 4-fold decrease of required fluorescence photons compared to single-photon MINFLUX. This suggests that detection with longer duration, less phototoxicity, or faster imaging speed can be warranted. Moreover, as different dyes can be excited simultaneously with two-photon excitation, 2p-MINFLUX have the potential for registration-free multicolor localizations, and simultaneous tracking of several fluorophores, which is essential for the study of molecular interactions. The application of higher-order Laguerre-Gaussian beams to MINFLUX is also discussed. We hope our study would promote new insights in optical vortexes and hollow beams for their application into MINFLUX and other microscopy techniques. In future experimental works, attention needs to be paid on choice of fluorophores, choice of wavelength of femtosecond laser, and attainable signal-to-background ratio.

\begin{acknowledgments}
\hspace*{\fill}

This work was financially supported by the National Natural Science Foundation of China (61729501, 31971376, 62025501), Beijing Natural Science Foundation (JQ18019), the National Key Research and Development Program of China (2017YFC0110202), and Shenzhen Science and Technology Program (KQTD20170810110913065).

Kun Zhao and Xinzhu Xu contributed equally to this work.
\end{acknowledgments}

\end{footnotesize}
\scriptsize
\bibliography{2pMINFLUX_ref.bib}

\begin{thebibliography}{28}%
\makeatletter
\providecommand \@ifxundefined [1]{%
 \@ifx{#1\undefined}
}%
\providecommand \@ifnum [1]{%
 \ifnum #1\expandafter \@firstoftwo
 \else \expandafter \@secondoftwo
 \fi
}%
\providecommand \@ifx [1]{%
 \ifx #1\expandafter \@firstoftwo
 \else \expandafter \@secondoftwo
 \fi
}%
\providecommand \natexlab [1]{#1}%
\providecommand \enquote  [1]{``#1''}%
\providecommand \bibnamefont  [1]{#1}%
\providecommand \bibfnamefont [1]{#1}%
\providecommand \citenamefont [1]{#1}%
\providecommand \href@noop [0]{\@secondoftwo}%
\providecommand \href [0]{\begingroup \@sanitize@url \@href}%
\providecommand \@href[1]{\@@startlink{#1}\@@href}%
\providecommand \@@href[1]{\endgroup#1\@@endlink}%
\providecommand \@sanitize@url [0]{\catcode `\\12\catcode `\$12\catcode
  `\&12\catcode `\#12\catcode `\^12\catcode `\_12\catcode `\%12\relax}%
\providecommand \@@startlink[1]{}%
\providecommand \@@endlink[0]{}%
\providecommand \url  [0]{\begingroup\@sanitize@url \@url }%
\providecommand \@url [1]{\endgroup\@href {#1}{\urlprefix }}%
\providecommand \urlprefix  [0]{URL }%
\providecommand \Eprint [0]{\href }%
\providecommand \doibase [0]{https://doi.org/}%
\providecommand \selectlanguage [0]{\@gobble}%
\providecommand \bibinfo  [0]{\@secondoftwo}%
\providecommand \bibfield  [0]{\@secondoftwo}%
\providecommand \translation [1]{[#1]}%
\providecommand \BibitemOpen [0]{}%
\providecommand \bibitemStop [0]{}%
\providecommand \bibitemNoStop [0]{.\EOS\space}%
\providecommand \EOS [0]{\spacefactor3000\relax}%
\providecommand \BibitemShut  [1]{\csname bibitem#1\endcsname}%
\let\auto@bib@innerbib\@empty
\bibitem [{\citenamefont {Balzarotti}\ \emph {et~al.}(2017)\citenamefont
  {Balzarotti}, \citenamefont {Eilers}, \citenamefont {Gwosch}, \citenamefont
  {Gynn{\aa}}, \citenamefont {Westphal}, \citenamefont {Stefani}, \citenamefont
  {Elf},\ and\ \citenamefont {Hell}}]{balzarotti2017nanometer}%
  \BibitemOpen
  \bibfield  {author} {\bibinfo {author} {\bibfnamefont {F.}~\bibnamefont
  {Balzarotti}}, \bibinfo {author} {\bibfnamefont {Y.}~\bibnamefont {Eilers}},
  \bibinfo {author} {\bibfnamefont {K.~C.}\ \bibnamefont {Gwosch}}, \bibinfo
  {author} {\bibfnamefont {A.~H.}\ \bibnamefont {Gynn{\aa}}}, \bibinfo {author}
  {\bibfnamefont {V.}~\bibnamefont {Westphal}}, \bibinfo {author}
  {\bibfnamefont {F.~D.}\ \bibnamefont {Stefani}}, \bibinfo {author}
  {\bibfnamefont {J.}~\bibnamefont {Elf}},\ and\ \bibinfo {author}
  {\bibfnamefont {S.~W.}\ \bibnamefont {Hell}},\ }\bibfield  {title} {\bibinfo
  {title} {Nanometer resolution imaging and tracking of fluorescent molecules
  with minimal photon fluxes},\ }\href@noop {} {\bibfield  {journal} {\bibinfo
  {journal} {Science}\ }\textbf {\bibinfo {volume} {355}},\ \bibinfo {pages}
  {606} (\bibinfo {year} {2017})}\BibitemShut {NoStop}%
\bibitem [{\citenamefont {Eilers}\ \emph {et~al.}(2018)\citenamefont {Eilers},
  \citenamefont {Ta}, \citenamefont {Gwosch}, \citenamefont {Balzarotti},\ and\
  \citenamefont {Hell}}]{eilers2018minflux}%
  \BibitemOpen
  \bibfield  {author} {\bibinfo {author} {\bibfnamefont {Y.}~\bibnamefont
  {Eilers}}, \bibinfo {author} {\bibfnamefont {H.}~\bibnamefont {Ta}}, \bibinfo
  {author} {\bibfnamefont {K.~C.}\ \bibnamefont {Gwosch}}, \bibinfo {author}
  {\bibfnamefont {F.}~\bibnamefont {Balzarotti}},\ and\ \bibinfo {author}
  {\bibfnamefont {S.~W.}\ \bibnamefont {Hell}},\ }\bibfield  {title} {\bibinfo
  {title} {Minflux monitors rapid molecular jumps with superior spatiotemporal
  resolution},\ }\href@noop {} {\bibfield  {journal} {\bibinfo  {journal}
  {Proceedings of the National Academy of Sciences}\ }\textbf {\bibinfo
  {volume} {115}},\ \bibinfo {pages} {6117} (\bibinfo {year}
  {2018})}\BibitemShut {NoStop}%
\bibitem [{\citenamefont {Pape}\ \emph {et~al.}(2020)\citenamefont {Pape},
  \citenamefont {Stephan}, \citenamefont {Balzarotti}, \citenamefont
  {B{\"u}chner}, \citenamefont {Lange}, \citenamefont {Riedel}, \citenamefont
  {Jakobs},\ and\ \citenamefont {Hell}}]{pape2020multicolor}%
  \BibitemOpen
  \bibfield  {author} {\bibinfo {author} {\bibfnamefont {J.~K.}\ \bibnamefont
  {Pape}}, \bibinfo {author} {\bibfnamefont {T.}~\bibnamefont {Stephan}},
  \bibinfo {author} {\bibfnamefont {F.}~\bibnamefont {Balzarotti}}, \bibinfo
  {author} {\bibfnamefont {R.}~\bibnamefont {B{\"u}chner}}, \bibinfo {author}
  {\bibfnamefont {F.}~\bibnamefont {Lange}}, \bibinfo {author} {\bibfnamefont
  {D.}~\bibnamefont {Riedel}}, \bibinfo {author} {\bibfnamefont
  {S.}~\bibnamefont {Jakobs}},\ and\ \bibinfo {author} {\bibfnamefont {S.~W.}\
  \bibnamefont {Hell}},\ }\bibfield  {title} {\bibinfo {title} {Multicolor 3d
  minflux nanoscopy of mitochondrial micos proteins},\ }\href@noop {}
  {\bibfield  {journal} {\bibinfo  {journal} {Proceedings of the National
  Academy of Sciences}\ }\textbf {\bibinfo {volume} {117}},\ \bibinfo {pages}
  {20607} (\bibinfo {year} {2020})}\BibitemShut {NoStop}%
\bibitem [{\citenamefont {Gwosch}\ \emph {et~al.}(2020)\citenamefont {Gwosch},
  \citenamefont {Pape}, \citenamefont {Balzarotti}, \citenamefont {Hoess},
  \citenamefont {Ellenberg}, \citenamefont {Ries},\ and\ \citenamefont
  {Hell}}]{gwosch2020minflux}%
  \BibitemOpen
  \bibfield  {author} {\bibinfo {author} {\bibfnamefont {K.~C.}\ \bibnamefont
  {Gwosch}}, \bibinfo {author} {\bibfnamefont {J.~K.}\ \bibnamefont {Pape}},
  \bibinfo {author} {\bibfnamefont {F.}~\bibnamefont {Balzarotti}}, \bibinfo
  {author} {\bibfnamefont {P.}~\bibnamefont {Hoess}}, \bibinfo {author}
  {\bibfnamefont {J.}~\bibnamefont {Ellenberg}}, \bibinfo {author}
  {\bibfnamefont {J.}~\bibnamefont {Ries}},\ and\ \bibinfo {author}
  {\bibfnamefont {S.~W.}\ \bibnamefont {Hell}},\ }\bibfield  {title} {\bibinfo
  {title} {Minflux nanoscopy delivers 3d multicolor nanometer resolution in
  cells},\ }\href@noop {} {\bibfield  {journal} {\bibinfo  {journal} {Nature
  Methods}\ }\textbf {\bibinfo {volume} {17}},\ \bibinfo {pages} {217}
  (\bibinfo {year} {2020})}\BibitemShut {NoStop}%
\bibitem [{\citenamefont {Hell}\ and\ \citenamefont
  {Wichmann}(1994)}]{hell1994breaking}%
  \BibitemOpen
  \bibfield  {author} {\bibinfo {author} {\bibfnamefont {S.~W.}\ \bibnamefont
  {Hell}}\ and\ \bibinfo {author} {\bibfnamefont {J.}~\bibnamefont
  {Wichmann}},\ }\bibfield  {title} {\bibinfo {title} {Breaking the diffraction
  resolution limit by stimulated emission: stimulated-emission-depletion
  fluorescence microscopy},\ }\href@noop {} {\bibfield  {journal} {\bibinfo
  {journal} {Optics Letters}\ }\textbf {\bibinfo {volume} {19}},\ \bibinfo
  {pages} {780} (\bibinfo {year} {1994})}\BibitemShut {NoStop}%
\bibitem [{\citenamefont {Betzig}\ \emph {et~al.}(2006)\citenamefont {Betzig},
  \citenamefont {Patterson}, \citenamefont {Sougrat}, \citenamefont
  {Lindwasser}, \citenamefont {Olenych}, \citenamefont {Bonifacino},
  \citenamefont {Davidson}, \citenamefont {Lippincott-Schwartz},\ and\
  \citenamefont {Hess}}]{betzig2006imaging}%
  \BibitemOpen
  \bibfield  {author} {\bibinfo {author} {\bibfnamefont {E.}~\bibnamefont
  {Betzig}}, \bibinfo {author} {\bibfnamefont {G.~H.}\ \bibnamefont
  {Patterson}}, \bibinfo {author} {\bibfnamefont {R.}~\bibnamefont {Sougrat}},
  \bibinfo {author} {\bibfnamefont {O.~W.}\ \bibnamefont {Lindwasser}},
  \bibinfo {author} {\bibfnamefont {S.}~\bibnamefont {Olenych}}, \bibinfo
  {author} {\bibfnamefont {J.~S.}\ \bibnamefont {Bonifacino}}, \bibinfo
  {author} {\bibfnamefont {M.~W.}\ \bibnamefont {Davidson}}, \bibinfo {author}
  {\bibfnamefont {J.}~\bibnamefont {Lippincott-Schwartz}},\ and\ \bibinfo
  {author} {\bibfnamefont {H.~F.}\ \bibnamefont {Hess}},\ }\bibfield  {title}
  {\bibinfo {title} {Imaging intracellular fluorescent proteins at nanometer
  resolution},\ }\href@noop {} {\bibfield  {journal} {\bibinfo  {journal}
  {Science}\ }\textbf {\bibinfo {volume} {313}},\ \bibinfo {pages} {1642}
  (\bibinfo {year} {2006})}\BibitemShut {NoStop}%
\bibitem [{\citenamefont {Rust}\ \emph {et~al.}(2006)\citenamefont {Rust},
  \citenamefont {Bates},\ and\ \citenamefont {Zhuang}}]{2006Stochastic}%
  \BibitemOpen
  \bibfield  {author} {\bibinfo {author} {\bibfnamefont {M.~J.}\ \bibnamefont
  {Rust}}, \bibinfo {author} {\bibfnamefont {M.}~\bibnamefont {Bates}},\ and\
  \bibinfo {author} {\bibfnamefont {X.}~\bibnamefont {Zhuang}},\ }\bibfield
  {title} {\bibinfo {title} {Stochastic optical reconstruction microscopy
  (storm) provides sub-diffraction-limit image resolution},\ }\href@noop {}
  {\bibfield  {journal} {\bibinfo  {journal} {Nature Methods}\ }\textbf
  {\bibinfo {volume} {3}},\ \bibinfo {pages} {793} (\bibinfo {year}
  {2006})}\BibitemShut {NoStop}%
\bibitem [{\citenamefont {Hell}\ \emph {et~al.}(1994)\citenamefont {Hell},
  \citenamefont {Stelzer}, \citenamefont {Lindek},\ and\ \citenamefont
  {Cremer}}]{hell1994confocal}%
  \BibitemOpen
  \bibfield  {author} {\bibinfo {author} {\bibfnamefont {S.~W.}\ \bibnamefont
  {Hell}}, \bibinfo {author} {\bibfnamefont {E.~H.}\ \bibnamefont {Stelzer}},
  \bibinfo {author} {\bibfnamefont {S.}~\bibnamefont {Lindek}},\ and\ \bibinfo
  {author} {\bibfnamefont {C.}~\bibnamefont {Cremer}},\ }\bibfield  {title}
  {\bibinfo {title} {Confocal microscopy with an increased detection aperture:
  type-b 4pi confocal microscopy},\ }\href@noop {} {\bibfield  {journal}
  {\bibinfo  {journal} {Optics Letters}\ }\textbf {\bibinfo {volume} {19}},\
  \bibinfo {pages} {222} (\bibinfo {year} {1994})}\BibitemShut {NoStop}%
\bibitem [{\citenamefont {Gustafsson}\ \emph {et~al.}(1999)\citenamefont
  {Gustafsson}, \citenamefont {Agard}, \citenamefont {Sedat} \emph
  {et~al.}}]{gustafsson1999i5m}%
  \BibitemOpen
  \bibfield  {author} {\bibinfo {author} {\bibfnamefont {M.~G.}\ \bibnamefont
  {Gustafsson}}, \bibinfo {author} {\bibfnamefont {D.}~\bibnamefont {Agard}},
  \bibinfo {author} {\bibfnamefont {J.}~\bibnamefont {Sedat}}, \emph {et~al.},\
  }\bibfield  {title} {\bibinfo {title} {I5m: 3d widefield light microscopy
  with better than 100nm axial resolution},\ }\href@noop {} {\bibfield
  {journal} {\bibinfo  {journal} {Journal of Microscopy}\ }\textbf {\bibinfo
  {volume} {195}},\ \bibinfo {pages} {10} (\bibinfo {year} {1999})}\BibitemShut
  {NoStop}%
\bibitem [{\citenamefont {Bewersdorf}\ \emph {et~al.}(2006)\citenamefont
  {Bewersdorf}, \citenamefont {Schmidt},\ and\ \citenamefont
  {Hell}}]{bewersdorf2006comparison}%
  \BibitemOpen
  \bibfield  {author} {\bibinfo {author} {\bibfnamefont {J.}~\bibnamefont
  {Bewersdorf}}, \bibinfo {author} {\bibfnamefont {R.}~\bibnamefont
  {Schmidt}},\ and\ \bibinfo {author} {\bibfnamefont {S.~W.}\ \bibnamefont
  {Hell}},\ }\bibfield  {title} {\bibinfo {title} {Comparison of i5m and
  4pi-microscopy},\ }\href@noop {} {\bibfield  {journal} {\bibinfo  {journal}
  {Journal of Microscopy}\ }\textbf {\bibinfo {volume} {222}},\ \bibinfo
  {pages} {105} (\bibinfo {year} {2006})}\BibitemShut {NoStop}%
\bibitem [{\citenamefont {Pavani}\ \emph {et~al.}(2009)\citenamefont {Pavani},
  \citenamefont {Thompson}, \citenamefont {Biteen}, \citenamefont {Lord},
  \citenamefont {Liu}, \citenamefont {Twieg}, \citenamefont {Piestun},\ and\
  \citenamefont {Moerner}}]{pavani2009three}%
  \BibitemOpen
  \bibfield  {author} {\bibinfo {author} {\bibfnamefont {S.~R.~P.}\
  \bibnamefont {Pavani}}, \bibinfo {author} {\bibfnamefont {M.~A.}\
  \bibnamefont {Thompson}}, \bibinfo {author} {\bibfnamefont {J.~S.}\
  \bibnamefont {Biteen}}, \bibinfo {author} {\bibfnamefont {S.~J.}\
  \bibnamefont {Lord}}, \bibinfo {author} {\bibfnamefont {N.}~\bibnamefont
  {Liu}}, \bibinfo {author} {\bibfnamefont {R.~J.}\ \bibnamefont {Twieg}},
  \bibinfo {author} {\bibfnamefont {R.}~\bibnamefont {Piestun}},\ and\ \bibinfo
  {author} {\bibfnamefont {W.}~\bibnamefont {Moerner}},\ }\bibfield  {title}
  {\bibinfo {title} {Three-dimensional, single-molecule fluorescence imaging
  beyond the diffraction limit by using a double-helix point spread function},\
  }\href@noop {} {\bibfield  {journal} {\bibinfo  {journal} {Proceedings of the
  National Academy of Sciences}\ }\textbf {\bibinfo {volume} {106}},\ \bibinfo
  {pages} {2995} (\bibinfo {year} {2009})}\BibitemShut {NoStop}%
\bibitem [{\citenamefont {Jia}\ \emph {et~al.}(2014)\citenamefont {Jia},
  \citenamefont {Vaughan},\ and\ \citenamefont {Zhuang}}]{jia2014isotropic}%
  \BibitemOpen
  \bibfield  {author} {\bibinfo {author} {\bibfnamefont {S.}~\bibnamefont
  {Jia}}, \bibinfo {author} {\bibfnamefont {J.~C.}\ \bibnamefont {Vaughan}},\
  and\ \bibinfo {author} {\bibfnamefont {X.}~\bibnamefont {Zhuang}},\
  }\bibfield  {title} {\bibinfo {title} {Isotropic three-dimensional
  super-resolution imaging with a self-bending point spread function},\
  }\href@noop {} {\bibfield  {journal} {\bibinfo  {journal} {Nature Photonics}\
  }\textbf {\bibinfo {volume} {8}},\ \bibinfo {pages} {302} (\bibinfo {year}
  {2014})}\BibitemShut {NoStop}%
\bibitem [{\citenamefont {Backlund}\ \emph {et~al.}(2018)\citenamefont
  {Backlund}, \citenamefont {Shechtman},\ and\ \citenamefont
  {Walsworth}}]{backlund2018fundamental}%
  \BibitemOpen
  \bibfield  {author} {\bibinfo {author} {\bibfnamefont {M.~P.}\ \bibnamefont
  {Backlund}}, \bibinfo {author} {\bibfnamefont {Y.}~\bibnamefont
  {Shechtman}},\ and\ \bibinfo {author} {\bibfnamefont {R.~L.}\ \bibnamefont
  {Walsworth}},\ }\bibfield  {title} {\bibinfo {title} {Fundamental precision
  bounds for three-dimensional optical localization microscopy with poisson
  statistics},\ }\href@noop {} {\bibfield  {journal} {\bibinfo  {journal}
  {Physical Review Letters}\ }\textbf {\bibinfo {volume} {121}},\ \bibinfo
  {pages} {023904} (\bibinfo {year} {2018})}\BibitemShut {NoStop}%
\bibitem [{\citenamefont {Xiao}\ and\ \citenamefont
  {Ha}(2017)}]{xiao2017flipping}%
  \BibitemOpen
  \bibfield  {author} {\bibinfo {author} {\bibfnamefont {J.}~\bibnamefont
  {Xiao}}\ and\ \bibinfo {author} {\bibfnamefont {T.}~\bibnamefont {Ha}},\
  }\bibfield  {title} {\bibinfo {title} {Flipping nanoscopy on its head},\
  }\href@noop {} {\bibfield  {journal} {\bibinfo  {journal} {Science}\ }\textbf
  {\bibinfo {volume} {355}},\ \bibinfo {pages} {582} (\bibinfo {year}
  {2017})}\BibitemShut {NoStop}%
\bibitem [{\citenamefont {Peticolas}\ \emph {et~al.}(1963)\citenamefont
  {Peticolas}, \citenamefont {Goldsborough},\ and\ \citenamefont
  {Rieckhoff}}]{peticolas1963double}%
  \BibitemOpen
  \bibfield  {author} {\bibinfo {author} {\bibfnamefont {W.~L.}\ \bibnamefont
  {Peticolas}}, \bibinfo {author} {\bibfnamefont {J.~P.}\ \bibnamefont
  {Goldsborough}},\ and\ \bibinfo {author} {\bibfnamefont {K.}~\bibnamefont
  {Rieckhoff}},\ }\bibfield  {title} {\bibinfo {title} {Double photon
  excitation in organic crystals},\ }\href@noop {} {\bibfield  {journal}
  {\bibinfo  {journal} {Physical Review Letters}\ }\textbf {\bibinfo {volume}
  {10}},\ \bibinfo {pages} {43} (\bibinfo {year} {1963})}\BibitemShut {NoStop}%
\bibitem [{\citenamefont {Sheppard}\ and\ \citenamefont
  {Gu}(1990)}]{sheppard1990image}%
  \BibitemOpen
  \bibfield  {author} {\bibinfo {author} {\bibfnamefont {C.~R.}\ \bibnamefont
  {Sheppard}}\ and\ \bibinfo {author} {\bibfnamefont {M.}~\bibnamefont {Gu}},\
  }\bibfield  {title} {\bibinfo {title} {Image formation in two-photon
  fluorescence microscopy},\ }\href@noop {} {\bibfield  {journal} {\bibinfo
  {journal} {Optik (Stuttgart)}\ }\textbf {\bibinfo {volume} {86}},\ \bibinfo
  {pages} {104} (\bibinfo {year} {1990})}\BibitemShut {NoStop}%
\bibitem [{\citenamefont {Denk}\ \emph {et~al.}(1990)\citenamefont {Denk},
  \citenamefont {Strickler},\ and\ \citenamefont {Webb}}]{denk1990two}%
  \BibitemOpen
  \bibfield  {author} {\bibinfo {author} {\bibfnamefont {W.}~\bibnamefont
  {Denk}}, \bibinfo {author} {\bibfnamefont {J.~H.}\ \bibnamefont
  {Strickler}},\ and\ \bibinfo {author} {\bibfnamefont {W.~W.}\ \bibnamefont
  {Webb}},\ }\bibfield  {title} {\bibinfo {title} {Two-photon laser scanning
  fluorescence microscopy},\ }\href@noop {} {\bibfield  {journal} {\bibinfo
  {journal} {Science}\ }\textbf {\bibinfo {volume} {248}},\ \bibinfo {pages}
  {73} (\bibinfo {year} {1990})}\BibitemShut {NoStop}%
\bibitem [{\citenamefont {Denk}\ \emph {et~al.}(2006)\citenamefont {Denk},
  \citenamefont {Piston},\ and\ \citenamefont {Webb}}]{denk2006multi}%
  \BibitemOpen
  \bibfield  {author} {\bibinfo {author} {\bibfnamefont {W.}~\bibnamefont
  {Denk}}, \bibinfo {author} {\bibfnamefont {D.~W.}\ \bibnamefont {Piston}},\
  and\ \bibinfo {author} {\bibfnamefont {W.~W.}\ \bibnamefont {Webb}},\
  }\bibfield  {title} {\bibinfo {title} {Multi-photon molecular excitation in
  laser-scanning microscopy},\ }in\ \href@noop {} {\emph {\bibinfo {booktitle}
  {Handbook of Biological Confocal Microscopy}}}\ (\bibinfo  {publisher}
  {Springer},\ \bibinfo {year} {2006})\ pp.\ \bibinfo {pages}
  {535--549}\BibitemShut {NoStop}%
\bibitem [{\citenamefont {Bestvater}\ \emph {et~al.}(2002)\citenamefont
  {Bestvater}, \citenamefont {Spiess}, \citenamefont {Stobrawa}, \citenamefont
  {Hacker}, \citenamefont {Feurer}, \citenamefont {Porwol}, \citenamefont
  {Berchner-Pfannschmidt}, \citenamefont {Wotzlaw},\ and\ \citenamefont
  {Acker}}]{bestvater2002two}%
  \BibitemOpen
  \bibfield  {author} {\bibinfo {author} {\bibfnamefont {F.}~\bibnamefont
  {Bestvater}}, \bibinfo {author} {\bibfnamefont {E.}~\bibnamefont {Spiess}},
  \bibinfo {author} {\bibfnamefont {G.}~\bibnamefont {Stobrawa}}, \bibinfo
  {author} {\bibfnamefont {M.}~\bibnamefont {Hacker}}, \bibinfo {author}
  {\bibfnamefont {T.}~\bibnamefont {Feurer}}, \bibinfo {author} {\bibfnamefont
  {T.}~\bibnamefont {Porwol}}, \bibinfo {author} {\bibfnamefont
  {U.}~\bibnamefont {Berchner-Pfannschmidt}}, \bibinfo {author} {\bibfnamefont
  {C.}~\bibnamefont {Wotzlaw}},\ and\ \bibinfo {author} {\bibfnamefont
  {H.}~\bibnamefont {Acker}},\ }\bibfield  {title} {\bibinfo {title}
  {Two-photon fluorescence absorption and emission spectra of dyes relevant for
  cell imaging},\ }\href@noop {} {\bibfield  {journal} {\bibinfo  {journal}
  {Journal of Microscopy}\ }\textbf {\bibinfo {volume} {208}},\ \bibinfo
  {pages} {108} (\bibinfo {year} {2002})}\BibitemShut {NoStop}%
\bibitem [{\citenamefont {M{\"u}tze}\ \emph {et~al.}(2012)\citenamefont
  {M{\"u}tze}, \citenamefont {Iyer}, \citenamefont {Macklin}, \citenamefont
  {Colonell}, \citenamefont {Karsh}, \citenamefont {Petr{\'a}{\v{s}}ek},
  \citenamefont {Schwille}, \citenamefont {Looger}, \citenamefont {Lavis},\
  and\ \citenamefont {Harris}}]{mutze2012excitation}%
  \BibitemOpen
  \bibfield  {author} {\bibinfo {author} {\bibfnamefont {J.}~\bibnamefont
  {M{\"u}tze}}, \bibinfo {author} {\bibfnamefont {V.}~\bibnamefont {Iyer}},
  \bibinfo {author} {\bibfnamefont {J.~J.}\ \bibnamefont {Macklin}}, \bibinfo
  {author} {\bibfnamefont {J.}~\bibnamefont {Colonell}}, \bibinfo {author}
  {\bibfnamefont {B.}~\bibnamefont {Karsh}}, \bibinfo {author} {\bibfnamefont
  {Z.}~\bibnamefont {Petr{\'a}{\v{s}}ek}}, \bibinfo {author} {\bibfnamefont
  {P.}~\bibnamefont {Schwille}}, \bibinfo {author} {\bibfnamefont {L.~L.}\
  \bibnamefont {Looger}}, \bibinfo {author} {\bibfnamefont {L.~D.}\
  \bibnamefont {Lavis}},\ and\ \bibinfo {author} {\bibfnamefont {T.~D.}\
  \bibnamefont {Harris}},\ }\bibfield  {title} {\bibinfo {title} {Excitation
  spectra and brightness optimization of two-photon excited probes},\
  }\href@noop {} {\bibfield  {journal} {\bibinfo  {journal} {Biophysical
  Journal}\ }\textbf {\bibinfo {volume} {102}},\ \bibinfo {pages} {934}
  (\bibinfo {year} {2012})}\BibitemShut {NoStop}%
\bibitem [{\citenamefont {Velasco}\ \emph {et~al.}(2015)\citenamefont
  {Velasco}, \citenamefont {Allgeyer}, \citenamefont {Yuan}, \citenamefont
  {Grutzendler},\ and\ \citenamefont {Bewersdorf}}]{velasco2015absolute}%
  \BibitemOpen
  \bibfield  {author} {\bibinfo {author} {\bibfnamefont {M.~G.~M.}\
  \bibnamefont {Velasco}}, \bibinfo {author} {\bibfnamefont {E.~S.}\
  \bibnamefont {Allgeyer}}, \bibinfo {author} {\bibfnamefont {P.}~\bibnamefont
  {Yuan}}, \bibinfo {author} {\bibfnamefont {J.}~\bibnamefont {Grutzendler}},\
  and\ \bibinfo {author} {\bibfnamefont {J.}~\bibnamefont {Bewersdorf}},\
  }\bibfield  {title} {\bibinfo {title} {Absolute two-photon excitation spectra
  of red and far-red fluorescent probes},\ }\href@noop {} {\bibfield  {journal}
  {\bibinfo  {journal} {Optics Letters}\ }\textbf {\bibinfo {volume} {40}},\
  \bibinfo {pages} {4915} (\bibinfo {year} {2015})}\BibitemShut {NoStop}%
\bibitem [{\citenamefont {Xu}\ and\ \citenamefont
  {Zipfel}(2015)}]{xu2015multiphoton}%
  \BibitemOpen
  \bibfield  {author} {\bibinfo {author} {\bibfnamefont {C.}~\bibnamefont
  {Xu}}\ and\ \bibinfo {author} {\bibfnamefont {W.~R.}\ \bibnamefont
  {Zipfel}},\ }\bibfield  {title} {\bibinfo {title} {Multiphoton excitation of
  fluorescent probes},\ }\href@noop {} {\bibfield  {journal} {\bibinfo
  {journal} {Cold Spring Harbor Protocols}\ }\textbf {\bibinfo {volume}
  {2015}},\ \bibinfo {pages} {pdb} (\bibinfo {year} {2015})}\BibitemShut
  {NoStop}%
\bibitem [{\citenamefont {Wolf}(1959)}]{wolf1959electromagnetic}%
  \BibitemOpen
  \bibfield  {author} {\bibinfo {author} {\bibfnamefont {E.}~\bibnamefont
  {Wolf}},\ }\bibfield  {title} {\bibinfo {title} {Electromagnetic diffraction
  in optical systems-i. an integral representation of the image field},\
  }\href@noop {} {\bibfield  {journal} {\bibinfo  {journal} {Proceedings of the
  Royal Society of London. Series A. Mathematical and Physical Sciences}\
  }\textbf {\bibinfo {volume} {253}},\ \bibinfo {pages} {349} (\bibinfo {year}
  {1959})}\BibitemShut {NoStop}%
\bibitem [{\citenamefont {Richards}\ and\ \citenamefont
  {Wolf}(1959)}]{richards1959electromagnetic}%
  \BibitemOpen
  \bibfield  {author} {\bibinfo {author} {\bibfnamefont {B.}~\bibnamefont
  {Richards}}\ and\ \bibinfo {author} {\bibfnamefont {E.}~\bibnamefont
  {Wolf}},\ }\bibfield  {title} {\bibinfo {title} {Electromagnetic diffraction
  in optical systems, ii. structure of the image field in an aplanatic
  system},\ }\href@noop {} {\bibfield  {journal} {\bibinfo  {journal}
  {Proceedings of the Royal Society of London. Series A. Mathematical and
  Physical Sciences}\ }\textbf {\bibinfo {volume} {253}},\ \bibinfo {pages}
  {358} (\bibinfo {year} {1959})}\BibitemShut {NoStop}%
\bibitem [{\citenamefont {Youngworth}\ and\ \citenamefont
  {Brown}(2000)}]{youngworth2000focusing}%
  \BibitemOpen
  \bibfield  {author} {\bibinfo {author} {\bibfnamefont {K.~S.}\ \bibnamefont
  {Youngworth}}\ and\ \bibinfo {author} {\bibfnamefont {T.~G.}\ \bibnamefont
  {Brown}},\ }\bibfield  {title} {\bibinfo {title} {Focusing of high numerical
  aperture cylindrical-vector beams},\ }\href@noop {} {\bibfield  {journal}
  {\bibinfo  {journal} {Optics Express}\ }\textbf {\bibinfo {volume} {7}},\
  \bibinfo {pages} {77} (\bibinfo {year} {2000})}\BibitemShut {NoStop}%
\bibitem [{\citenamefont {Zhan}(2006)}]{zhan2006properties}%
  \BibitemOpen
  \bibfield  {author} {\bibinfo {author} {\bibfnamefont {Q.}~\bibnamefont
  {Zhan}},\ }\bibfield  {title} {\bibinfo {title} {Properties of circularly
  polarized vortex beams},\ }\href@noop {} {\bibfield  {journal} {\bibinfo
  {journal} {Optics Letters}\ }\textbf {\bibinfo {volume} {31}},\ \bibinfo
  {pages} {867} (\bibinfo {year} {2006})}\BibitemShut {NoStop}%
\bibitem [{\citenamefont {Zhao}\ \emph {et~al.}(2007)\citenamefont {Zhao},
  \citenamefont {Edgar}, \citenamefont {Jeffries}, \citenamefont {McGloin},\
  and\ \citenamefont {Chiu}}]{zhao2007spin}%
  \BibitemOpen
  \bibfield  {author} {\bibinfo {author} {\bibfnamefont {Y.}~\bibnamefont
  {Zhao}}, \bibinfo {author} {\bibfnamefont {J.~S.}\ \bibnamefont {Edgar}},
  \bibinfo {author} {\bibfnamefont {G.~D.}\ \bibnamefont {Jeffries}}, \bibinfo
  {author} {\bibfnamefont {D.}~\bibnamefont {McGloin}},\ and\ \bibinfo {author}
  {\bibfnamefont {D.~T.}\ \bibnamefont {Chiu}},\ }\bibfield  {title} {\bibinfo
  {title} {Spin-to-orbital angular momentum conversion in a strongly focused
  optical beam},\ }\href@noop {} {\bibfield  {journal} {\bibinfo  {journal}
  {Physical Review Letters}\ }\textbf {\bibinfo {volume} {99}},\ \bibinfo
  {pages} {073901} (\bibinfo {year} {2007})}\BibitemShut {NoStop}%
\bibitem [{\citenamefont {Xie}\ \emph {et~al.}(2013)\citenamefont {Xie},
  \citenamefont {Liu}, \citenamefont {Jin}, \citenamefont {Santangelo},\ and\
  \citenamefont {Xi}}]{xie2013analytical}%
  \BibitemOpen
  \bibfield  {author} {\bibinfo {author} {\bibfnamefont {H.}~\bibnamefont
  {Xie}}, \bibinfo {author} {\bibfnamefont {Y.}~\bibnamefont {Liu}}, \bibinfo
  {author} {\bibfnamefont {D.}~\bibnamefont {Jin}}, \bibinfo {author}
  {\bibfnamefont {P.~J.}\ \bibnamefont {Santangelo}},\ and\ \bibinfo {author}
  {\bibfnamefont {P.}~\bibnamefont {Xi}},\ }\bibfield  {title} {\bibinfo
  {title} {Analytical description of high-aperture sted resolution with
  0--2$\pi$ vortex phase modulation},\ }\href@noop {} {\bibfield  {journal}
  {\bibinfo  {journal} {JOSA A}\ }\textbf {\bibinfo {volume} {30}},\ \bibinfo
  {pages} {1640} (\bibinfo {year} {2013})}\BibitemShut {NoStop}%
\end{thebibliography}%


%

\end{document}